\documentclass[aps,prd,twocolumn,preprintnumbers,superscriptaddress,%
nofootinbib,amsmath,amssymb]{revtex4}

\begin{document}

\preprint{arXiv:0909.1347 [hep-th]}

\title{Lifshitz Black Hole in Three Dimensions}

\author{Eloy Ay\'on--Beato}
\email{ayon-beato-at-fis.cinvestav.mx}
\affiliation{Departamento~de~F\'{\i}sica,~CINVESTAV--IPN,~Apdo.%
~Postal~14--740,~07000,~M\'exico~D.F.,~M\'exico.}
\author{Alan Garbarz}
\email{alan-at-df.uba.ar}
\affiliation{Departamento de F\'{\i}sica, Universidad de Buenos Aires
FCEN - UBA, Ciudad Universitaria, Pabell\'on 1, 1428, Buenos Aires,
Argentina.}
\author{Gaston Giribet}
\email{gaston-at-df.uba.ar}
\affiliation{Center for Cosmology and Particle Physics,
at the Physics Department of New York University CCPP - NYU,
4 Washington Place NY10003, USA.}
\author{Mokhtar Hassa\"{\i}ne}
\email{hassaine-at-inst-mat.utalca.cl}
\affiliation{Instituto de Matem\'atica y F\'{\i}sica,
Universidad de Talca, Casilla 747, Talca, Chile.}

\begin{abstract}
We show that three-dimensional massive gravity admits Lifshitz metrics with
generic values of the dynamical exponent $z$ as exact
solutions. At the point $z=3$, exact black hole solutions which
are asymptotically Lifshitz arise. These spacetimes are
three-dimensional analogues of those that were recently
proposed as gravity duals for anisotropic scale invariant fixed
points.
\end{abstract}

\maketitle

The enormous success of gauge-gravity duality \cite{Malda} has
triggered the interest in generalizing the holographic
techniques to other areas of physics. Recently, the attempts to
generalize AdS/CFT correspondence to non-relativistic condensed
matter physics have received considerable attention. Besides
being an active line of research, this has given raise to very
interesting new ideas; see Ref.~\cite{Hartnoll} and references
therein for a review.

Recently, candidates to be gravity duals for non-relativistic
scale invariant theories, both exhibiting Galilean invariance
or not, have been proposed. In Refs.~\cite{Son,McGreevy},
spacetimes whose isometry group is the so-called
Schr\"{o}dinger group were proposed to be gravity duals for
Galilean and scale invariant systems. In Ref.~\cite{Kachru},
the scale invariant fixed points that do not exhibit Galilean
symmetry were also analyzed, and the metric of the
corresponding gravity duals were introduced (see
Eq.~(\ref{ansatz0}) below). These metrics manifestly exhibit
the anisotropic scale invariance
\begin{equation}
t\mapsto \lambda^{z}t, \qquad \vec{x}\mapsto \lambda\vec{x},
\label{A}
\end{equation}
which is characterized by the dynamical critical exponent $z$.
The value $z=1$ corresponds to the standard scaling behavior
of conformal invariant systems.

The typical example of a model with such a scaling symmetry is
the Lifshitz model with $z=2$; this is the reason why the
metrics of the proposed gravity duals are usually referred to
as Lifshitz spacetimes. This type of Lifshitz fixed point
appears in systems of strongly correlated electrons and other
interesting problems in condensed matter physics. In turn,
having a holographic description for these phenomena would be
of great importance to describe condensed matter systems in the
strongly coupled regime.

The basic idea is to look for spacetime geometries that
incarnate the dynamical scaling symmetry (\ref{A}). The
spacetimes that were identified as possible gravity duals for
these Lifshitz fixed points are \cite{Kachru}
\begin{equation}
ds^2=-\frac{r^{2z}}{l^{2z}}dt^2+\frac{l^2}{r^2}dr^2
+\frac{r^2}{l^2}d\vec{x}^2.\label{ansatz0}
\end{equation}
where $\vec{x}$ is a $d$-dimensional vector. It is simple to
see that these spacetimes are invariant under the rescaling
$(t,\vec{x},r)\mapsto(\lambda^{z}t,\lambda\vec{x},\lambda^{-1}r)$.
It was also shown in \cite{Kachru} that metrics (\ref{ansatz0})
with $d=2$ arise as solutions of General Relativity with a
negative cosmological constant and $p$-form gauge fields as
sources.

The proposal for the holographic prescription in \cite{Kachru}
follows the standard AdS/CFT recipe. In the ($d+2$)-dimensional
bulk defined by the spacetime (\ref{ansatz0}), one considers a
massive scalar mode whose asymptotic behavior in the near
boundary limit takes the form
$\varphi(r)\simeq\varphi_0^{(-)}r^{-\Delta^{(z)}_-}
+\varphi_0^{(+)}r^{-\Delta^{(z)}_+}$. This yields the relation
between the mass of the scalar mode ${\text m}$ in the bulk and
the conformal dimension $\Delta^{(z)}$ of the associated
operator ${\mathcal O}_{\Delta^{(z)}}$ in the boundary theory,
given by $\Delta^{(z)}_{\pm } ( \Delta^{(z)}_{\pm }-z-d )
={\text m}^{2}l^{2}$. As consequence, the
Breitenlohner-Freedman type bound $\text{m}^2l^2>-(z+d)^{2}/4$
arises. Then, one considers the branch $\Delta_-^{(z)}$, whose
fall-off does not spoil the asymptotic behavior of the metric
at infinity. Finally, one is ready to match bulk and boundary
observables computing correlators of physical operators. In
\cite{Kachru}, the case of a two-point correlation function
$\left< {\mathcal O}_{\Delta^{(2)}}(x){\mathcal
O}_{\Delta^{(2)}}(y) \right> $ was analyzed.

Of great importance is to introduce finite temperature effects
in the story. With this motivation, black hole solutions in
Lifshitz spaces (\ref{ansatz0}) were also investigated in the
literature. However, in spite of the apparently simple form of
the spacetimes (\ref{ansatz0}), the problem of finding analytic
exact black hole solutions which asymptote these metrics turned
out to be a highly non-trivial problem. In Ref.~\cite{Mann}, a
particular solution was found, which corresponds to a
four-dimensional topological black hole which is asymptotically
Lifshitz with dynamical exponent $z=2$. In
Refs.~\cite{BertoldiI,BertoldiII}, numerical solutions were
also explored. Lifshitz black holes were also studied in
Refs.~\cite{Marika,DanielssonI,DanielssonII}, and while this letter
was being finished the paper \cite{McGreevy2} just appeared,
where a similar analytical solution was found for $z=2$ with
$d=2$. The problem of embedding these black holes in string
theory was addressed in Ref.~\cite{TakayanagiI}, where a
remarkable solution with $z=3/2$ was found. Moreover, from the
analysis performed in Ref.~\cite{TakayanagiI}, it becomes
evident how difficult is generalizing the solution to other
values of $z$. In particular, some no-go theorems for the
string theory embedding have been discussed in
Ref.~\cite{TakayanagiII}.

The main result of this letter is to show the existence of
black hole solutions of three-dimensional massive gravity that
are asymptotically Lifshitz with $z=3$. In addition, we will
also establish that the vacua of three-dimensional massive
gravity includes metrics (\ref{ansatz0}) for generic values of
dynamical critical exponent $z$.

The theory we will consider is the so-called New Massive
Gravity (NMG) \cite{NMG}, which has attracted much attention
recently due to its very appealing properties. NMG is defined
by supplementing Einstein-Hilbert action with the particular
square-curvature terms which gives raise to field equations
with a second order trace. At the linearized level, it is
equivalent to the Fierz-Pauli action for a massive spin-2
particle in three dimensions, which turns out to be a unitary
model. The space of solutions of the theory was studied
recently, and it became clear that it includes geometries of
great interest, like black holes, warped-AdS$_3$ spaces, and
AdS-waves.

The action of the NMG is \cite{NMG}
\begin{equation}
S=\frac{1}{16\pi G}\int d^{3}x\sqrt{-g}\left[ R-2\lambda
-\frac{1}{m^{2}} \left( R_{\mu \nu }R^{\mu \nu
}-\frac{3}{8}R^{2}\right) \right] . \label{eq:S}
\end{equation}
The associated field equations read
\begin{equation}
R_{\mu \nu }-\frac{1}{2} R g_{\mu \nu }+\lambda {g}_{\mu \nu
}-\frac{1}{2m^{2}}K_{\mu \nu }=0,  \label{eq:NMG}
\end{equation}
where
\begin{eqnarray}
K_{\mu \nu } &=&2\square {R}_{\mu\nu}-\frac{1}{2}\nabla_{\mu}
\nabla_{\nu }{R}-\frac{1}{2}\square {R}g_{\mu\nu}
+4R_{\mu\alpha\nu\beta}R^{\alpha\beta}  \nonumber \\
&&{}-\frac{3}{2}RR_{\mu\nu}-R_{\alpha\beta}R^{\alpha\beta}g_{\mu\nu}
+\frac{3}{8}R^{2}g_{\mu \nu }.  \label{eom}
\end{eqnarray}
Here, we will consider $G=1/8$. It is also convenient to define
the dimensionless parameters
\[
m^{2}l^{2}=y, \quad \lambda l^{2}=w.
\]
This gravity theory exhibits special properties at the points
\begin{equation}
y=\pm 1/2. \label{SpPo}
\end{equation}
In particular, at $y=+1/2$, the central charge associated to
the Virasoro algebra that generates the group of asymptotic
AdS$_{3}$ symmetries vanishes. At this point, solutions with
interesting properties have been exhibited in
\cite{AdSWaves,Clement} where the asymptotically AdS$_3$
solutions were shown to present a relaxed fall-off at infinity
\cite{ponjasI,ponjasII}. We will see below that $y=+1/2$ is
precisely the point where the dynamical exponent $z$ of the
solutions we find takes the value $z=1$. In this case, the
Lifshitz spacetime becomes AdS$_{3}$ as it can be seen also
from the scaling property (\ref{A}) which corresponds to that
of the conformal group.

At $y=-1/2$, the theory also exhibits special properties. It
was shown in \cite{AdSWaves} that at this point the scalar
modes of gravitational waves in AdS$_{3}$ space precisely
saturates the Breitenlonher-Freedman bound (BF); this fact is
closely related to the emergence of solutions with a relaxed
fall-off, which are typically given by logarithmic asymptotic
branches in convenient system of coordinates. Also for
$y=-1/2$, the theory admits interesting black hole solutions
\cite{Troncoso,NMGII} that generalize the static BTZ black hole
\cite{BTZ,BHTZ}. These (former) black holes also present a
weakened version of Brown-Henneaux AdS$_{3}$ boundary
conditions \cite{Brown:1986nw}. As we will show below, the
dynamical exponent of our Lifshitz vacua when $y=-1/2$
corresponds to $z=3$, and this is precisely the point where we
will also find exact analytic black hole solutions with
Lifshitz asymptotic. For completeness, let us mention that NMG
admits solutions with full or partial Schr\"{o}dinger isometry
\cite{AdSWaves}. These solutions correspond to
three-dimensional analogues of the spacetimes considered in
\cite{Son,McGreevy} as gravity duals for cold atoms. The full
Schr\"{o}dinger group arises at the point $y=17/2$.

In this letter, we are concerned with spacetimes with no
Galilean symmetry. We first analyze solutions that are of the
Lifshitz-type \cite{Kachru}. It is possible to verify that the
equations of motion (\ref{eom}) admit solutions of the form
(\ref{ansatz0}) for a generic dynamical exponent $z$. In fact,
it parameterizes the mass and the cosmological constant as
\begin{equation}
y=-\frac{1}{2}(z^{2}-3z+1),\quad w=-\frac{1}{2}(z^{2}+z+1).
\label{curves}
\end{equation}
In other words, Lifshitz vacua exist for generic $z$ provided
an appropriate tuning of the coupling constants. This implies
that Lifshitz solutions with $z\neq1$ are only possible if the
coupling constants satisfy the conditions $\lambda
l^{2}\le-3/8$ and $m^{2}l^{2}\le5/8$. On the other hand, we
also have the AdS$_{3}$ vacua $z=1$, and its identifications,
for a different relation between the coupling constants
$w=-[1+1/(4y)]$. Curvature invariants $R$ and $R_{\mu \nu }R^{\mu \nu }$
associated to solution
(\ref{ansatz0}) do depend on $z$; these are constant and yield
the relation $g^{\mu \nu} K_{\mu \nu} = R_{\mu \nu }R^{\mu \nu } - 
\frac{3}{8}R^2=2m^2\lambda $.

Let us turn to the problem of finding black hole solutions of
NMG that are asymptotically Lifshitz. In order to accomplish
this goal, we consider the following ansatz
\[
ds^{2}=-\frac{r^{2z}}{l^{2z}}F(r)dt^{2}+\frac{l^{2}}{r^{2}}H(r)dr^{2}+\frac{%
r^{2}}{l^{2}}dx^{2}
\]
where $F(r)$ and $H(r)$ are functions of the radial coordinate.
We will demand these functions to obey
$\lim_{r\rightarrow\infty}F(r)=
\lim_{r\rightarrow\infty}H^{-1}(r)=1$ and to present a
single-zero at a given radius $r=r_{+}$ where the horizon would
be located, namely $F(r_{+})=H^{-1}(r_{+})=0$.

Intriguingly, for $z=3$, which corresponds to the particular
point $y=-1/2$, $w=-13/2$, the field equations with the
appropriate asymptotic conditions turn out to be solved by
\begin{equation}
F(r)=H^{-1}(r)=1-\frac{Ml^2}{r^{2}}, \label{eq:F(r)}
\end{equation}
where $M$ is an integration constant. Then, the static
asymptotically Lifshitz black hole for $z=3$ is given by
\begin{equation}
ds^{2}=-\frac{r^6}{l^6}\left(1-\frac{Ml^2}{r^2}\right)dt^{2}
+\frac{dr^2}{\left(\frac{r^2}{l^2}-M\right)}+r^2d\phi^2,
\label{eq:sLbh}
\end{equation}
where we have renamed $\phi=x/l$ and the identification
$\phi=\phi+2\pi$ have been considered.  The metric
(\ref{eq:sLbh}) presents a curvature singularity at $r=0$ and a
single event horizon located at 
\begin{equation}
r_{+}=l\sqrt{M} .
\end{equation}
It is interesting to note that the $z=3$ Lifshitz scale
symmetry, $t\mapsto\lambda^{3}t$, $x\mapsto\lambda x$,
$r\mapsto\lambda^{-1}r$, is preserved provided the parameter is
allowed to rescale as $M\mapsto\lambda^{-2}M$.

Solution (\ref{eq:sLbh}) presents curvature singularity at $r=0$.
In fact, the curvature invariants diverge at the origin; namely 
\begin{equation*}
R = -\frac{26}{l^2}+\frac{8M}{r^2} , \ \ 
R_{\mu \nu }R^{\mu \nu }= \frac{260}{l^4} - \frac{152M}{l^2r^2} + 
\frac{24M^2}{r^4} .
\end{equation*}
The Hawking temperature associated to the black hole solution
(\ref{eq:sLbh}) can be easily computed by requiring regularity
on the tip of the Euclidean geometry after periodic
identification. This yields the result
\begin{equation}
T_H = \frac{r_+^3}{2\pi l^4} =  \frac{M^{3/2}}{2\pi l} ,
\end{equation}
which is consistent with the behavior $T_H \sim r_+^{z}$ found
in other examples. A complete analysis of the thermodynamical
properties would require a better understanding of the
computation of conserved charges for these backgrounds in NMG
and, in particular, of the counterterms involved. This will be
discussed elsewhere \cite{csoon}.

It turns out that the radial configuration for a massive scalar
field $\varphi(r)$ in the black hole background (\ref{eq:sLbh})
can be explicitly found in terms of hypergeometric functions.
Namely,
\begin{eqnarray*}
\varphi(r)&=&\varphi_0^{(-)}\,
F\left(k,1+k;1+2k;\frac{Ml^2}{r^2}\right)r^{-2-2k}\\
&+&\varphi_0^{(+)} \,
F\left(-k,1-k;1-2k;\frac{Ml^2}{r^2}\right)r^{-2+2k}.
\end{eqnarray*}
where $k=-\frac{1}{2}\sqrt{\text{m}^2l^2+4}$ and $\text{m}$ is
the mass of the scalar field. Notice that this exhibits the
expected behavior at large distances, which reproduces the
falling of $\sim r^{-\Delta^{(3)}_{\pm}}$ discussed above, as
$\Delta^{(3)}_{\pm }=2\pm 2k$. It is also worth noticing that
the solution exhibits a logarithmic dependence at $k=0$, namely
$\text{m}^2l^2=-4$, where a BF-type bound is saturated. This is
a common feature in similar solutions on AdS space.

The solution for the field configuration 
$\varphi(t,r,\phi)= \Phi(r) e^{ik\phi -i\omega t}$, that also depends on $t$ 
and $\phi $, can be also explicitly found. In this case, the radial dependence $\Phi (r)$
can be expressed in terms of HeunC special functions, which are solutions to the Heun's
Confluent differential equation.

Coefficients $\varphi^{(\pm )}_0$ in the expression above are ultimately determined by imposing
boundary conditions at the horizon; see the analysis of
\cite{McGreevy2}. Actually, analyzing the similarities with the
case $z=2$, $d=2$ of \cite{McGreevy2} results interesting. As
in \cite{McGreevy2}, the scalar field equation in the
background (\ref{eq:sLbh}) admits an exact expression in terms
of hypergeometric functions. The resemblance with the case
$z=2$, $d=2$ would make possible to adapt the analysis of
\cite{McGreevy2} to our example and arrive to similar
conclusions, like the exclusion of ultralocal form for the dual
correlators.

Of course, a black hole solution for the case $z=1$ is already
known; it corresponds to the static BTZ black hole
\cite{BTZ,BHTZ}
\begin{equation}
ds^{2}=-\frac{r^{2}}{l^{2}}\left( 1-\frac{Ml^{2}}{r^{2}}\right) dt^{2}+\frac{%
dr^{2}}{\left( \frac{r^{2}}{l^{2}}-M\right) }+r^{2}d\phi ^{2}.
\label{BTZ}
\end{equation}
In some sense, the solution (\ref{eq:sLbh}) can be thought of
as a cousin of the three-dimensional static BTZ black hole,
which arises for a different value of the dynamical exponent
$z$. Within the parametrization (\ref{curves}) the two black
hole solutions $z=1$ and $z=3$ appear at the two special points
(\ref{SpPo}) discussed above.

It is worth noticing that the black hole solution
(\ref{eq:sLbh}) is conformally equivalent to a black string
solution of the form
\begin{equation}
ds^{2}=\frac{r^{2}}{l^{2}}\left( - f(r)dt^{2}+\frac{dr^{2}}{f(r)}%
+dx^{2}\right) .  \label{BString}
\end{equation}
with $f(r)=\frac{r^{4}}{l^{4}}\left(
1-\frac{Ml^{2}}{r^{2}}\right) $, and where the coordinate
$x=l\phi$ is again uncompactified by taking the universal
covering of $\phi$. It can be seen that, by boosting this black
string solution along the $x$ direction, one can construct a
rotating version of the spacetime (\ref{eq:sLbh}). This
spinning version of the solution will be discussed somewhere
else \cite{csoon}.

The fact of having found that the theory defined by action
(\ref{eq:S}) admits Lifshitz solutions with generic values for
the dynamical exponent $z$ is highly non-trivial. To illustrate
how difficult finding such a solutions in a higher-curvature
theory can be, let us consider the case of five-dimensional
Einstein-Gauss-Bonnet theory
\begin{eqnarray}
S &=&\frac{1}{2\pi}\int d^{5}x\sqrt{-g}\left[ R+12l^{-2}-\xi
l^{2}\left( R^{2}-4R_{\mu \nu }R^{\mu \nu }\right. \right. +  \nonumber \\
&&\left. \left. R_{\mu \nu \alpha \beta }R^{\mu \nu \alpha \beta
}\right) \right],  \label{CS}
\end{eqnarray}
which is a simpler example as it yields field equations of
second order. It can be shown that, for generic values of the
coupling constant $\xi$, this theory only admits solutions of
the Lifshitz type for $z=1$; that is, the only solution
corresponds to locally AdS$_{5}$.\footnote{G.G.\ is grateful to
Mat\'{\i}as Aiello for a discussion about this point.} The only
point where curious features arise is for $\xi=1/4$, where the
theory suffers enhancement of symmetry. Indeed, for $\xi=1/4$,
the action (\ref{CS}) can be written as a five-dimensional
Chern-Simons action, and thus it enjoys local gauge invariance
under $SO(2,4)$ group. At this point, Lifshitz solutions are
admitted for generic values of $z$ due to the well-known
degeneracies of the Chern-Simons theory \cite{TroncosoJulioDotti}; see also \cite{Pang}
for a consideration of the $ISO(1,4)$ invariant Chern-Simons
theory $1/\xi = 0$. This degeneracy in the value of the
dynamical exponent is closely related to the
non-renormalization of $z$ that happens at $\xi =1/4$, which
was implicitly observed in \cite{Maloney}. Nevertheless, even
at this special point where degeneracy in $z$ arises,
asymptotically Lifshitz black holes do not seem to exist (for
$z\neq 1$). This makes the solution (\ref{eq:sLbh}) of
particular interest.


Besides its uses as a remarkably simple model to explore the
$d>1$ analogues, it could be interesting to investigate whether
a more direct application of this $d=1$ Lifshitz solution to
condensed matter physics exists. In particular, exploring the
relation to the models discussed in \cite{Yu} would be
interesting.

Finally, one might also wonder whether a wider sector of Lifshitz
fixed points arises if one considers the generalization of three-dimensional
massive gravity that amounts to including the Cotton tensor in the gravity
action \cite{DJTTMG,NMG}. After all, one knows that, in certain cases, the
inclusion of the Cotton tensor produces the enlargement of the space of
allowed configurations \cite{AdSWaves}. However, one can verify that
this is not the case of the Lifshitz solutions. Remarkably, the
addition of the Cotton tensor in the action generically excludes the
Lifshitz configurations, as the only cases that are allowed correspond
to $z=0$ and $z=1$.

\begin{acknowledgments}
The authors thank M.~Aiello, J.~Oliva and R.~Troncoso for
conversations. The work of A.G. is supported by University of
Buenos Aires. G.G. is Member of the CIC-CONICET, Argentina, on
leave of absence from University of Buenos Aires. This work has
been partially supported by grant 1090368 from FONDECYT, by the
project Redes de Anillos R04 from CONICYT, by grant UBACyT X861
from UBA, by grant PICT 00849 from ANPCyT, and by grants 82443
and 45946-F from CONACyT.
\end{acknowledgments}


\end{document}